\documentclass[letterpaper,american,reprint, aps,superscriptaddress]{revtex4-1}
\usepackage[T1]{fontenc}
\usepackage[latin9]{inputenc}
\usepackage{amsmath}
\usepackage{amssymb}
\usepackage{graphicx}

\makeatletter

\pdfpageheight\paperheight
\pdfpagewidth\paperwidth

\makeatother

\usepackage{babel}
\begin{document}
\global\long\def\degc{\;^{\circ}\text{C}}%
\global\long\def\od{\text{OD}}%

\title{Single-photon hologram of a zero-area pulse}
\author{Micha\l{} Lipka}
\email{m.lipka@cent.uw.edu.pl}

\affiliation{Centre for Quantum Optical Technologies, Centre of New Technologies,
~\\
 University of Warsaw, Banacha 2c, 02-097 Warsaw, Poland }
\author{Micha\l{} Parniak}
\email{m.parniak@cent.uw.edu.pl}

\affiliation{Centre for Quantum Optical Technologies, Centre of New Technologies,
~\\
 University of Warsaw, Banacha 2c, 02-097 Warsaw, Poland }
\affiliation{Niels Bohr Institute, University of Copanhagen, Blegdamsvej 17, 2100
Copenhagen, Denmark.}
\date{\today}
\begin{abstract}
Single photons exhibit inherently quantum and unintuitive properties
such as the Hong-Ou-Mandel effect, demonstrating their bosonic and
quantized nature, yet at the same time may corresponds to single excitations
of spatial or temporal modes with a very complex structure. Those
two features are rarely seen together. Here we experimentally demonstrate
how the Hong-Ou-Mandel effect can be spectrally-resolved and harnessed
to characterize a complex temporal mode of a single-photon \textendash{}
a zero-area pulse \textendash{} obtained via a resonant interaction
of a THz-bandwidth photon with a narow GHz-wide atomic transition
of atomic vapor. The combination of bosonic quantum behavior with
bandwidth-mismatched light-atom interaction is of fundamental importance
for deeper understanding of both phenomena, as well as their engineering
offering applications in characterization of ultrafast transient processes.
\end{abstract}
\maketitle
Single photons (SPs) exhibit a plethora of highly nonclassical features
manifesting their quantized and bosonic nature and demonstrating the
meanders of quantum theory. One of classic examples is the Hong-Ou-Mandel
(HOM) effect \citep{Hong1987}. When two identical photons enter two
respective input ports of a balanced beamsplitter (BS), the photons
always leave together via a single output port. In consequence no
coincidences between the ports can be observed. This unintuitive feature
fundamentally stems from the destructive interference of two-photon
amplitudes corresponding to two scenarios each with one of the photons
reflected and one transmitted. Interestingly, observing reminiscent
coincidences with a single-photon camera (i.e. spatially/angularly
resolved) allows one to probe and localize the wavefront differences
of the two photons, which has been leveraged to precisely measure
a single-photon wavefront in a method reminiscent of classical holography
\citep{Chrapkiewicz2016}. Two-photon interferograms, measured with
either spatial/angular or temporal/spectral resolution, are also at
the core of super-resolution imaging \citep{Parniak2018}, quantum
fingerprinting \citep{Jachura2017,jachura_visibility-based_2018,Lipka2020b}
and photon-pair source characterization \citep{Thiel2020,Jin2015,Prakash2021,Montaut2018}.
Spectrally-resolved HOM effect has been measured with a dispersive
fiber spectrometer \citep{Gerrits2015,Orre2019}. HOM interference
extends beyond the photonic realm and has been shown for other bosonic
(quasi)particles such as atoms \citep{Kaufman2014,Lopes2015}, phonons
\citep{Toyoda2015} or spin-waves \citep{Li2016,Parniak2019}.

The non-classical features of SPs become intriguing when the spatial
or temporal mode has a nontrivial structure with added qualitative
features such as orbital angular momentum (OAM). In the spatial domain,
mode structuring has lead to insights and applications such as remote
object identification \citep{UribePatarroyo2013}, improved sensitivity
\citep{Fickler2012}, and uncertainty relations for OAM \citep{Leach2010}.
However, with mostly single mode optical architectures, the mode engineering
in time-frequency (TF) degree-of-freedom (DoF) attracts most attention,
enabling bandwidth matching \citep{Karpinski2017}, generating and
manipulating high-dimensional entanglement on-chip \citep{Kues2017}
and studies of global-versus-local two-photon interference (TPI) in
quantum networks \citep{Nitsche2020}.

SPs exhibit complex light-matter interactions, mostly studied in the
classical regime and leading to complex modal structure. One such
a case is the resonant interaction of an ultrafast pulse with a slowly
relaxing medium e.g. a femtosecond pulse passing through atomic vapor
\textendash{} a case demonstrated to produce zero-area (ZA) or $0\pi$
pulses \citep{McCall1967,Crisp1970} with temporal envelopes consisting
of alternating $\pm$ sign lobes. This interaction has been demonstrated
for SP states and Rb vapor \citep{Costanzo2016}, exploring a vastly
unharnessed region of light-matter interaction between a THz-bandwidth
photon and a GHz-wide (Doppler broadened) atomic transition. While
the photon is rarely absorbed, the interaction is highly dispersive
leaving the photon in the ZA temporal shape and imprinting a spectral
phase (SPHI). Costanzo \emph{et al}. \citep{Costanzo2016} characterized
SP ZA pulses in the temporal domain via homodyne detection, which
is a robust tomography method, yet never provides optimal information
due to inherent shot noise. Single-photon holography can directly
reconstruct the SPHI resulting from bandwidth-mismatched (BM) light-matter
interaction, without local oscillator (LO) optimization and the shot
noise contribution inherent to homodyne techniques. Characterizing
the SPHI at a single-photon level is of fundamental importance for
quantum coherent control techniques \citep{Meshulach1998,Devolder2021}
and utilization of the TF DoF in quantum networks \citep{Nitsche2020}.
High-bandwidth ultrafast photons can also efficiently interact with
atomic vapors via two-photon transitions \citep{Carvalho2020,Meshulach1998}.
Furthermore, ultrafast frequency combs have been used to probe temporal
dynamics of two-photon transitions via direct frequency comb spectroscopy
\citep{Marian2004} combining high temporal resolution and a vast
spectral range.
\begin{figure}[b]
\includegraphics[width=1\columnwidth]{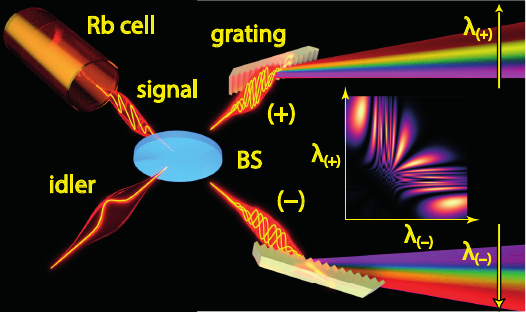}\caption{\label{fig:idea} Spectral single-photon holography of an ultrafast
zero-area (ZA) photon. Starting from a pair of identical broadband
(10 nm, 100 fs) single photons(signal, idler), one (signal) interacts
with hot $\mathrm{^{87}Rb}$ vapor, forming a ZA temporal shape and
acquiring a spectral phase (SPHI). Signal and idler photons are interfered
on a balanced beamsplitter (BS) which output $\pm$ modes are spectrally
resolved via diffraction gratings. Coincidence detection between wavelength
$\lambda_{\pm}$ components shows a footprint of the ZA single-photon's
SPHI \textendash{} a manifest of the HOM effect. (inset) Simulated
map of spectrally-resolved ($\lambda_{\pm}$ ) coincidences \textendash{}
an analogue of classical interferogram.}
\end{figure}

In this Letter we combine two highly nonintuitive concepts and demonstrate
spectral single-photon holography (SSPH) applied to the characterization
of ultrafast ZA SP pulses. Idea of our experiment is depicted in Fig.
\ref{fig:idea}. We employ spontaneous parametric down-conversion
(SPDC) to produce pairs of SPs \textendash{} an idler and a signal
\textendash{} with central wavelengths of $795\ \mathrm{nm}$. The
signal photon with $10\ \mathrm{nm}$ bandwidth passes through a hot
Rb vapor, resonantly interacting with the D1-line electronic transition,
and obtaining both a ZA temporal shape and intrinsically a nontrivial
SPHI profile imprinted through light-matter interaction. The photon
pair is then interfered on a BS. The output ports ($\pm)$ are analyzed
with a single-photon spectrometer and coincidences in wavelength coordinates
($\lambda_{\pm}$) are counted. The inset of Fig. \ref{fig:idea}
depicts a simulated coincidence map. In the absence of the BS and
the Rb cell the coincidence map would just be the joint spectral intensity
(JSI) of the two-photon state. Notably, if only the Rb was removed,
we would ideally see no coincidences at all, indicating the SPHIs
of the two-photons are identical. A coincidence pattern uniquely corresponds
to a SPHI difference between the photons and carries a footprint of
the interaction between a THz-bandwidth photon and a GHz atomic transition.

Fundamentally, our demonstration combines a vastly unexplored regime
of bandwidth-mismatched light-matter interaction with a purely quantum
effect of TPI. The previous SP ZA pulses demonstration \citep{Costanzo2016}
employed temporal homodyne detection requiring LO temporal mode optimization
\citep{Polycarpou2012}. In comparison, SSPH  does not require LO
or any optimization, working readily for any kind of SPHI profiles.
Importantly for quantum metrology, our method also avoids the shot
noise inherent to homodyning.

We envisage applications of SSPH for characterization of ultrafast
transient phenomena such as chemical reactions or for biological measurements
{[}see supplementary material (SM) \cite{SM}{]}, which could benefit from negligible \nocite{nesmeyanov_gary_1963}\nocite{Kallmann1999}\nocite{Kolenderski2009}\nocite{Karout2007}\nocite{Bone1986}\nocite{Costanzo2016}\nocite{Kay1993}\nocite{Sekatski2012}\nocite{Schleich2001}\nocite{GarrisonChiao}\nocite{Henriksson2009}\nocite{Tang2019}\nocite{Jacques2013}absorption of the probe photons. Ultrafast pulses have proved indispensable
where time resolution is required e.g. for time-resolved photoemission
tomography of molecular orbitals \citep{Wallauer2021}, probing the
transition states of chemical reactions \citep{ROSKER1988,Dantus1987}
or their coherent control \citep{Potter1992,Brumer1992}. SSPH can
supplement these methods as recent times see more and more proposals
for using quantum light \citep{Mukamel2020,Dorfman2016} and TPI effects
in spectroscopy \citep{Dorfman2021}. For instance, femtosecond transition
state spectroscopy \citep{ROSKER1988,Rose1988,Dantus1987} relies
on a pair of pump and delayed probe femtosecond pulses. SSPH could
replace the typical probe fluorescence signal for a direct noninvasive
characterization. SP operation also promises flexibility in molecular
control \citep{Brown2006} via quantum non-demolition continuous measurements
\citep{Meng2020} and relying on the TPI and coincidence post-selection,
provides robustness to the noise from scattering surroundings of the
biological samples. While being noninvasive, the SSPH remains an ultrafast
and interferometric scheme able to replace probe signals in different
spectroscopic methods \citep{Klinteberg2005,Hori2006}. The advantage
of noninvasive SP probing can be quantified with a Fisher information
per damage to the sample \citep{Wolfgramm2013}\emph{.}

\begin{figure}[h]
\includegraphics[width=1\columnwidth]{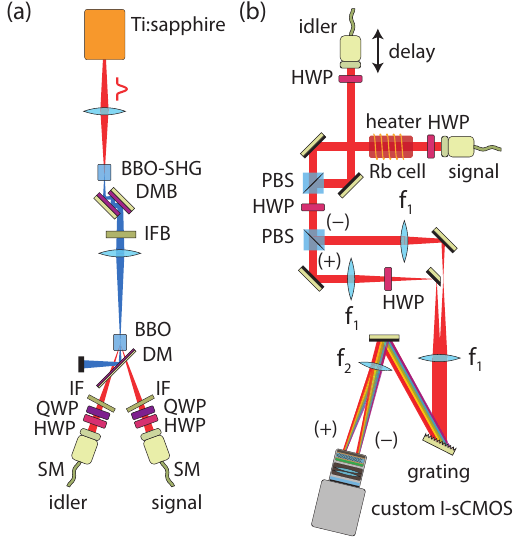}\caption{\label{fig:setup} (a) Source of single-photon pairs. $100\;\mathrm{fs}$
pulses from Ti:Sapphire laser (central wavelength $795\;\mathrm{nm}$)
are frequency-doubled in a BBO crystal (BBO-SHG) and used to pump
type-I SPDC in the second crystal (BBO). Pairs of photons are spectrally
filtered (IF, see main text) and coupled to single-mode polarization-maintaining
(PM) fibers (SM) which select highest-correlated transverse modes
from the SPDC emission cone. Dichroic mirrors (DM for $\approx800\;\mathrm{nm}$
, DMB for $\approx400\;\mathrm{nm}$) and a $400\;\mathrm{nm}$ bandpass
($40\;\mathrm{nm}$ FWHM) inteference filter (IFB - IF for $400\;\mathrm{nm}$)
separate pumping beams. Quarter-wave (QWP) and half-wave (HWP) plates
allow for polarization matching to the PM fiber. (b) Setup for SSPH.
Signal photon passes through heated Rb cell, interacting resonantly
with D1 Rb line and obtaining a SPHI profile. The output $\pm$ modes
of the interferometer are spatially separated on another PBS. Imaging
setup (focal length $f_{1}=150\;\mathrm{mm}$) superimposes the $\pm$
modes spatially on a diffraction grating while separating them angularly.
The last HWP in the $(+)$ mode path rotates the polarization to be
perpendicular to the grating grooves ensuring maximal efficiency.
The grating is far-field imaged (focal length $f_{2}=300\;\mathrm{mm}$)
onto an intensified single-photon\textendash sensitive camera (I-sCMOS),
spatially separating the spectral components of $\pm$ modes into
distinct camera frame regions ($\pm)$.}
\end{figure}
Our experimental setup consists of a SP source (Fig. \ref{fig:setup}(a))
and of SSPH part (Fig. \ref{fig:setup}(b)). To produce ultrashort
SP states we employ the type-I noncollinear SPDC process in a beta
barium borate (BBO, $2\ \mathrm{mm}$ length along optical axis),
pumped with a focused pump beam with 397.5 nm central wavelength,
Gaussian beam radius of $w_{0}=70\;\mu\text{m}$ and $100\ \mathrm{mW}$
average power. To obtain the blue pump, we employ second harmonic
generation of femtosecond pulses (100 fs, central wavelength of 795
nm) from a Ti:Sapphire laser (Spectra-Physics MaiTai), in a second
BBO crystal ($0.5\ \mathrm{mm}$ length along optical axis). The SPDC
emission is spectrally filtered with an interference filter (FWHM
$10\ \mathrm{nm}$) tilted to have the central transmission wavelength
at $796.7\;\mathrm{nm}$. Two Gaussian transverse modes are selected
by coupling to single-mode fibers. The modes are chosen for highest
correlation and SP brightness. Selected modes, corresponding to signal
and idler photons, enter the second part of the setup. Signal photons
pass through a heated Rb vapor cell. The cell temperature controls
the optical depth {[}OD, see SM{]} which determines the strength of
light-matter coupling. The photon leaves the Rb cell in an ZA temporal
shape and with its SPHI given by a single Lorentzian resonance profile
centered on the $\lambda_{0}=795\ \mathrm{nm}$ D1 Rb line \citep{Kallmann1999}:
\begin{equation}
\varphi_{s}(\lambda)=\od\times\frac{x(\lambda)}{1+x(\lambda)^{2}},\label{eq:philam}
\end{equation}
 where $x(\lambda)=2\pi\tau c(\lambda-\lambda_{0})/\lambda_{0}^{2}$
with $c$ denoting the speed of light and $\tau$ is the Doppler-broadened
lifetime of the excited state ranging from $215\ \mathrm{ps}$ to
$240\ \mathrm{ps}$ for employed Rb temperatures (see SM). The idler
photon is delayed to match the signal (the linear component of the
SPHI is compensated). The photon pair is then interfered on a polarization-based
equivalent of a balanced BS, a setup employed previously in Ref. \citep{Chrapkiewicz2016}.
Initially, the signal and idler photons have their polarizations rotated
to vertical and horizontal, respectively, which allows superimposing
their spatial modes on a polarizing beamsplitter (PBS). The photons'
polarizations are then jointly rotated on a half-wave plate (HWP),
to diagonal and anti-diagonal, respectively. Finally, the photons
interfere and are spatially separated by another PBS which output
ports correspond to the $\pm$ outputs and are far-field imaged onto
a mirror and a D-shaped mirror, respectively, allowing to separate
the ports angularly, while imaging both on a single diffraction grating
($\mathrm{1200\ {lines/mm}}$, $750\ \mathrm{nm}$ blaze). The grating
is far-field imaged on an ultrafast intensified CMOS camera \citep{Lipka2021,Lipka2018}.
On the camera frame the $\pm$ ports appear as $140\times5\;\text{px}$
regions, with the longer dimension corresponding to $\lambda_{\pm}$.
We collect $8.2\times10^{4}$ camera frames per second with an average
of $\bar{n}\approx0.2$ photons per frame ($1.4\times10^{-4}$ per
pixel). The camera pixels are not photon-number-resolving; however,
on average we expect two or more photons to be misclassified as a
SP only once per $7.2\times10^{4}$ frames (see SM).

Let us denote by $0\leq n(\lambda_{\pm})\leq5$ the number of photons
registered at the $\lambda_{\pm}$ coordinates in a single frame (one
photon per pixel, 5 px per spectral point), and by $\langle.\rangle$
the average over collected frames. A raw coincidence map $\mathcal{R}(\lambda_{+},\lambda_{-})=\langle n(\lambda_{+})n(\lambda_{-})\rangle$
is a normalized histogram of events where a photon pair is registered
in a single frame with the first (second) photon in $+$ ($-$) region
at the $\lambda_{+}$ ($\lambda_{-}$) coordinate. We subtract accidental
coincidences $\mathcal{A}(\lambda_{+},\lambda_{-})=\langle n(\lambda_{+})\rangle\langle n(\lambda_{-})\rangle$
to obtain the coincidence map:
\begin{equation}
\mathcal{C}(\lambda_{+},\lambda_{-})=\mathcal{R}(\lambda_{+},\lambda_{-})-\mathcal{A}(\lambda_{+},\lambda_{-}),
\end{equation}
which is the photon-number covariance. The subtraction is required
due to many experiment repetitions per single camera frame, creating
artificial coincidences (see SM). To predict the form of $\mathcal{C}(\lambda_{+},\lambda_{-})$
we consider the two-photon component of the signal ($s$) and idler
($i$) joint wavefunction $\Psi(\lambda_{s},\lambda_{i})$. The probability
of observing a coincidence at spectral coordinates $\lambda_{\pm}$
of $\pm$ BS ports is given by:
\begin{equation}
P_{\text{c}}(\lambda_{+},\lambda_{-})=\frac{1}{4}|\Psi(\lambda_{+},\lambda_{-})-\Psi(\lambda_{-},\lambda_{+})|^{2}.
\end{equation}
We assume that the photons are identical except for the SPHI $\varphi_{s}(\lambda)$
of the signal mode i.e. $\Psi(\lambda_{-},\lambda_{+})=\Psi(\lambda_{+},\lambda_{-})\exp\{i[\varphi_{s}(\lambda_{-})-\varphi_{s}(\lambda_{+})]\}$.
Hence,
\begin{equation}
P_{\text{c }}(\lambda_{+},\lambda_{-})=\frac{1}{2}[1-\cos(\varphi_{s}(\lambda_{+})-\varphi_{s}(\lambda_{-}))]\times|\Psi(\lambda_{+},\lambda_{-})|^{2},\label{eq:pcoinc_cos}
\end{equation}
where the first term lets us reconstruct the SPHI $\varphi_{s}(\lambda)$
and the second corresponds to the JSI of the photon pair. Additional
mode mismatch between the photons decreases the interference visibility,
corresponding to $\cos(\ldots)\rightarrow\mathcal{V}\cos(\ldots)$
for visibility $\mathcal{V}\leq1$.

\begin{figure}[h]
\includegraphics[width=1\columnwidth]{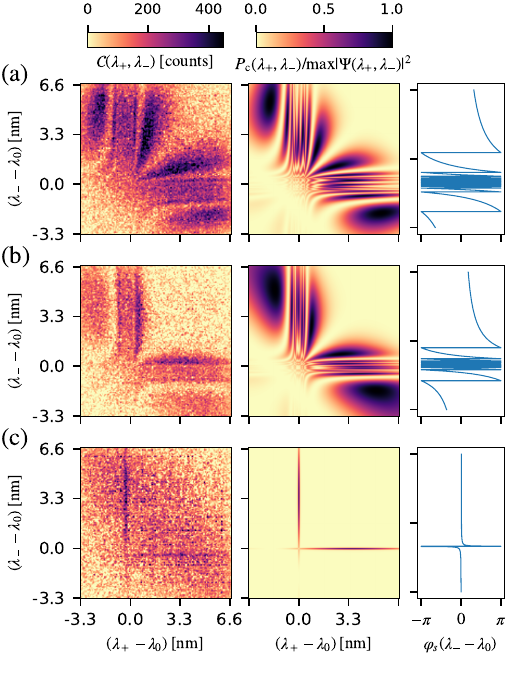}\caption{\label{fig:panels} Spectral single-photon holograms of an ultrafast
$100\;\mathrm{fs}$ photon resonantly interacting with Rb vapor heated
to (a) $T_{1}=188\protect\degc$, (b) $T_{2}=174\protect\degc$, (c)
$T_{3}=86\protect\degc$. Left column represents experimental results
i.e. the observed photon number covariance $\mathcal{C}(\lambda_{+},\lambda_{-})$
(coincidences with subtracted background) in spectral coordinates
$\lambda_{\pm}-\lambda_{0}$ between $\pm$ ports of a BS. The BS
interferes the measured photon with the reference. Central column
corresponds to a theoretical prediction of a spectrally-resolved coincidence
probability $P_{c}(\lambda_{+},\lambda_{-})$ with no imperfections
($\mathcal{V}=1$), as given by Eq. (\ref{eq:pcoinc_cos}). Right
column: theoretical prediction for the SPHI $\varphi_{s}(\lambda_{-}-\lambda_{0})$
modulo $2\pi$, given by Eq. (\ref{eq:philam}).}
\end{figure}
The coincidence maps for three Rb cell temperatures ($T_{1}=188\degc$,
$T_{2}=174\degc$, $T_{3}=86\degc$) are presented in Fig. \ref{fig:panels}
along with theoretical predictions. The fidelity $0\leq\mathcal{F}\leq1$
\citep{Jozsa1994} between the experimental and theoretical maps normalized
to a unit sum, yields $94\%$, $86\%$, $89\%$ for $T_{1},T_{2}$and
$T_{3}$, respectively. In all maps the coincidences are most strongly
present on a broad stripe along the anti-diagonal. This feature stems
from JSI $|\Psi(\lambda_{+},\lambda_{-})|^{2}$ of the SPDC photon
pairs, which in our case are spectrally correlated. See SM for comparison
of simulated interferograms with correlated and uncorrelated photons
and JSI maps. The characteristic cross at $\lambda_{\pm}=795\;\text{nm}$
corresponds to the Rb resonance where the phase variation becomes
too rapid to be resolved. The coincidences in this region correspond
to a phase-averaged case. The phase sign flip around $\lambda_{0}$,
illustrated in the right column of Fig. \ref{fig:panels}, follows
from Eq. (\ref{eq:philam}). While the brief interaction between a
SP and Rb atoms has the most pronounced spectral footprint at higher
temperatures, corresponding to fitted optical depths of $\od(T_{1})=4.6\times10^{3}$
and $\od(T_{2})=2.6\times10^{3}$, the presence of much cooler Rb
at $T_{3}$ is still distinctly identifiable, despite a comparably
low $\od(T_{3})\approx20$ yielding a peak-to-peak SPHI variation
of 20 rad. This observation suggests metrological applications in
non-disturbing sensing of the sample presence. The OD values corresponding
to SPHIs reconstructed from the experimental data are in a good agreement
with the theory prediction for independently measured Rb cell temperatures.
Even without prior knowledge of the phase profile $\varphi(\lambda)$
standard holographic reconstruction techniques \citep{Bone1986,Karout2007}
can extract $\varphi(\lambda)$ since the problem is analogous to
the processing of classical interferograms. We have verified the Fourier-domain
retrieval of $\varphi_{s}(\lambda_{+})-\varphi_{s}(\lambda_{-})$
modulo $2\pi$, for the case of $T_{1}=188\degc$ (see SM).

The interference visibility in SSPH is not only a benchmark of the
setup quality, but estimated locally $\mathcal{V}(\lambda_{+},\lambda_{-})$
carries unique information. The total phase $\varphi_{s}(\lambda_{+})-\varphi_{s}(\lambda_{-})$
estimated from the coincidence map is a 2-dimensional abundant representation
of a 1-dimensional $\varphi_{s}(\lambda)$ enhancing estimation of
$\varphi_{s}(\lambda)$ and guaranteeing that the $\cos$ term in
Eq. (\ref{eq:pcoinc_cos}) cannot be constant over the spectral range
of $\varphi_{s}(\lambda)$ unless equal to unity (in which case $P_{c}=0$).
Hence, when the local visibility $\mathcal{V}(\lambda_{+},\lambda_{-})$
falls to 0 over extended regions with coincidences present $P_{c}(\lambda_{+},\lambda_{-})\neq0$,
it is a strong indicator of rapid oscillations below resolution (averaging
$\cos$ to 0). The method can thus detect \emph{presence} of spectral
features on sub-resolution scales. Finally, we have quantified the
visibility by dividing a smoothened version of the experimental coincidence
map into square regions (1 nm sidelength) and locally estimating the
visibility from the maximal and minimal value within the region. This
way we obtained average visibility of $\mathcal{V\approx}0.69\pm0.16$,
$\mathcal{V\approx}0.79\pm0.12$ and $\mathcal{V\approx}0.88\pm0.09$
for $T_{1}=188\degc$, $T_{2}=174\degc$ and $T_{3}=86\degc$, respectively,
where the uncertainties correspond to one standard deviation across
the regions. For more details see SM. For comparison, using classical
light the maximal attainable visibility of TPI is 50\%.

\paragraph*{Metrological advantage of SSPH over homodyne detection. }

In conventional homodyning, we measure a variance of homodyne current
of the signal SP. To gain full information, we consider a multi-pixel
measurement and analyze covariance (see SM for setup examples), which
has been previously considered in the time domain \citep{Qin2015}.
The homodyne signal is zero-mean with covariance of $\sim[\frac{{1}}{2}\delta(\lambda_{+}-\lambda_{-})+|\psi_{s}(\lambda_{+})\psi_{s}(\lambda_{-})|\cos(\varphi_{s}(\lambda_{+})-\varphi_{s}(\lambda_{-}))]$
with an inherent shot noise component (first term) and $\psi_{s}(\lambda)$
being the SP wavefunction, which is in contrast to SSPH (Eq. \ref{eq:pcoinc_cos}).
To further elucidate the advantage, consider a more direct scenario
of self-guided tomography \citep{Ferrie2014}, where spectro-temporal
shaping is employed both for a LO and the reference SP. We strive
to shape the reference to be the same as the signal photon. The homodyne
case involves maximizing the the inherently noisy variance. The SSPH
involves minimization of the coincidence count, which becomes a low-noise
signal with no offset. This can be directly demonstrated by considering
the estimation of residual distinguishability $\alpha\ll1$ between
signal and reference modes, as derived in the SM in the context of
Fisher information $F_{\alpha}$ per experimental shot. We obtain
$F_{\alpha}\sim\alpha^{-1/2}$ for the homodyning and $F_{\alpha}\sim\alpha^{-1}$
for the SSPH, which unequivocally demonstrates the preferable scaling
of the SSPH. With no prior knowledge $(\alpha\approx1)$ hybrid of
homodyning followed by SSPH may be beneficial.

In this Letter we experimentally combined highly unintuitive phenomena
bringing together the bosonic and quantum nature of SP light and the
complex spectral structure of ultrafast ZA pulses obtained in resonant
BM interaction between a SP and $^{87}$Rb vapors. Our experiment
demonstrates how the Hong-Ou-Mandel effect can be spectrally resolved
and harnessed for characterization of ultrafast SPs in a holography-like
method. The broadband ultrafast photons carry a unique spectral footprint
of the resonant interaction with narrow atomic transitions, corresponding
to their ZA temporal shape. Demonstrated herein, SSPH both supplements
and extends previous homodyne measurements of SP ZA pulses, as well
as promises unique applications in probing ultrafast transient phenomena
such as picosecond-scale chemical reactions. Notably, no matching
of the reference to the photon is required. This may on one hand come
as a surprise, since visibility of interference is reduced as the
two photons cease to match in the time domain. Our method solves this
via spectrally-resolved detection, where local visibility of spectral
HOM interference is high. Furthermore, it broadens the fundamental
understanding of SP BM light-atom interactions and brings closer the
prospect of engineering those phenomena for the range of applications.
Apart from spectroscopic applications, such interactions may find
natural applications in quantum information processing in the spectral
domain \citep{Brecht2015,Lukens2017,KhodadadKashi2021}.
\begin{acknowledgments}
This scientific work has been funded by Polish science budget funds
for years 2019 to 2023 as a research project within the ``Diamentowy
Grant'' programme of the Ministry of Education and Science (DI2018
010848), by the Foundation for Polish Science (MAB/2018/4 \textquotedblleft Quantum
Optical Technologies\textquotedblright ) and by the Office of Naval
Research (N62909-19-1-2127). The \textquotedbl Quantum Optical Technologies\textquotedblright{}
project is carried out within the International Research Agendas programme
of the Foundation for Polish Science co-financed by the European Union
under the European Regional Development Fund. We would like to thank
M. Jachura and W. Wasilewski for fruitful discussions and K. Banaszek
for the generous support.
\end{acknowledgments}

\bibliographystyle{apsrev}
\bibliography{shom}

\begin{thebibliography}{66}
\expandafter\ifx\csname natexlab\endcsname\relax\def\natexlab#1{#1}\fi
\expandafter\ifx\csname bibnamefont\endcsname\relax
  \def\bibnamefont#1{#1}\fi
\expandafter\ifx\csname bibfnamefont\endcsname\relax
  \def\bibfnamefont#1{#1}\fi
\expandafter\ifx\csname citenamefont\endcsname\relax
  \def\citenamefont#1{#1}\fi
\expandafter\ifx\csname url\endcsname\relax
  \def\url#1{\texttt{#1}}\fi
\expandafter\ifx\csname urlprefix\endcsname\relax\def\urlprefix{URL }\fi
\providecommand{\bibinfo}[2]{#2}
\providecommand{\eprint}[2][]{\url{#2}}

\bibitem[{\citenamefont{Hong et~al.}(1987)\citenamefont{Hong, Ou, and
  Mandel}}]{Hong1987}
\bibinfo{author}{\bibfnamefont{C.~K.} \bibnamefont{Hong}},
  \bibinfo{author}{\bibfnamefont{Z.~Y.} \bibnamefont{Ou}}, \bibnamefont{and}
  \bibinfo{author}{\bibfnamefont{L.}~\bibnamefont{Mandel}},
  \bibinfo{journal}{Phys. Rev. Lett.} \textbf{\bibinfo{volume}{59}},
  \bibinfo{pages}{2044} (\bibinfo{year}{1987}).

\bibitem[{\citenamefont{Chrapkiewicz et~al.}(2016)\citenamefont{Chrapkiewicz,
  Jachura, Banaszek, and Wasilewski}}]{Chrapkiewicz2016}
\bibinfo{author}{\bibfnamefont{R.}~\bibnamefont{Chrapkiewicz}},
  \bibinfo{author}{\bibfnamefont{M.}~\bibnamefont{Jachura}},
  \bibinfo{author}{\bibfnamefont{K.}~\bibnamefont{Banaszek}}, \bibnamefont{and}
  \bibinfo{author}{\bibfnamefont{W.}~\bibnamefont{Wasilewski}},
  \bibinfo{journal}{Nat. Photonics} \textbf{\bibinfo{volume}{10}},
  \bibinfo{pages}{576} (\bibinfo{year}{2016}).

\bibitem[{\citenamefont{Parniak et~al.}(2018)\citenamefont{Parniak,
  Bor{\'{o}}wka, Boroszko, Wasilewski, Banaszek, and
  Demkowicz-Dobrza{\'{n}}ski}}]{Parniak2018}
\bibinfo{author}{\bibfnamefont{M.}~\bibnamefont{Parniak}},
  \bibinfo{author}{\bibfnamefont{S.}~\bibnamefont{Bor{\'{o}}wka}},
  \bibinfo{author}{\bibfnamefont{K.}~\bibnamefont{Boroszko}},
  \bibinfo{author}{\bibfnamefont{W.}~\bibnamefont{Wasilewski}},
  \bibinfo{author}{\bibfnamefont{K.}~\bibnamefont{Banaszek}}, \bibnamefont{and}
  \bibinfo{author}{\bibfnamefont{R.}~\bibnamefont{Demkowicz-Dobrza{\'{n}}ski}},
  \bibinfo{journal}{Phys. Rev. Lett.} \textbf{\bibinfo{volume}{121}},
  \bibinfo{pages}{250503} (\bibinfo{year}{2018}).

\bibitem[{\citenamefont{Jachura et~al.}(2017)\citenamefont{Jachura, Lipka,
  Jarzyna, and Banaszek}}]{Jachura2017}
\bibinfo{author}{\bibfnamefont{M.}~\bibnamefont{Jachura}},
  \bibinfo{author}{\bibfnamefont{M.}~\bibnamefont{Lipka}},
  \bibinfo{author}{\bibfnamefont{M.}~\bibnamefont{Jarzyna}}, \bibnamefont{and}
  \bibinfo{author}{\bibfnamefont{K.}~\bibnamefont{Banaszek}},
  \bibinfo{journal}{Opt. Express} \textbf{\bibinfo{volume}{25}},
  \bibinfo{pages}{27475} (\bibinfo{year}{2017}).

\bibitem[{\citenamefont{Jachura et~al.}(2018)\citenamefont{Jachura, Jarzyna,
  Lipka, Wasilewski, and Banaszek}}]{jachura_visibility-based_2018}
\bibinfo{author}{\bibfnamefont{M.}~\bibnamefont{Jachura}},
  \bibinfo{author}{\bibfnamefont{M.}~\bibnamefont{Jarzyna}},
  \bibinfo{author}{\bibfnamefont{M.}~\bibnamefont{Lipka}},
  \bibinfo{author}{\bibfnamefont{W.}~\bibnamefont{Wasilewski}},
  \bibnamefont{and} \bibinfo{author}{\bibfnamefont{K.}~\bibnamefont{Banaszek}},
  \bibinfo{journal}{Phys. Rev. Lett.} \textbf{\bibinfo{volume}{120}},
  \bibinfo{pages}{110502} (\bibinfo{year}{2018}).

\bibitem[{\citenamefont{{Lipka} et~al.}(2020)\citenamefont{{Lipka}, {Jarzyna},
  and {Banaszek}}}]{Lipka2020b}
\bibinfo{author}{\bibfnamefont{M.}~\bibnamefont{{Lipka}}},
  \bibinfo{author}{\bibfnamefont{M.}~\bibnamefont{{Jarzyna}}},
  \bibnamefont{and}
  \bibinfo{author}{\bibfnamefont{K.}~\bibnamefont{{Banaszek}}},
  \bibinfo{journal}{IEEE J. Sel. Areas Commun.} \textbf{\bibinfo{volume}{38}},
  \bibinfo{pages}{496} (\bibinfo{year}{2020}).

\bibitem[{\citenamefont{Thiel et~al.}(2020)\citenamefont{Thiel, Davis, Sun,
  D'Ornellas, Jin, and Smith}}]{Thiel2020}
\bibinfo{author}{\bibfnamefont{V.}~\bibnamefont{Thiel}},
  \bibinfo{author}{\bibfnamefont{A.~O.~C.} \bibnamefont{Davis}},
  \bibinfo{author}{\bibfnamefont{K.}~\bibnamefont{Sun}},
  \bibinfo{author}{\bibfnamefont{P.}~\bibnamefont{D'Ornellas}},
  \bibinfo{author}{\bibfnamefont{X.-M.} \bibnamefont{Jin}}, \bibnamefont{and}
  \bibinfo{author}{\bibfnamefont{B.~J.} \bibnamefont{Smith}},
  \bibinfo{journal}{Opt. Express} \textbf{\bibinfo{volume}{28}},
  \bibinfo{pages}{19315} (\bibinfo{year}{2020}).

\bibitem[{\citenamefont{Jin et~al.}(2015)\citenamefont{Jin, Gerrits, Fujiwara,
  Wakabayashi, Yamashita, Miki, Terai, Shimizu, Takeoka, and Sasaki}}]{Jin2015}
\bibinfo{author}{\bibfnamefont{R.-B.} \bibnamefont{Jin}},
  \bibinfo{author}{\bibfnamefont{T.}~\bibnamefont{Gerrits}},
  \bibinfo{author}{\bibfnamefont{M.}~\bibnamefont{Fujiwara}},
  \bibinfo{author}{\bibfnamefont{R.}~\bibnamefont{Wakabayashi}},
  \bibinfo{author}{\bibfnamefont{T.}~\bibnamefont{Yamashita}},
  \bibinfo{author}{\bibfnamefont{S.}~\bibnamefont{Miki}},
  \bibinfo{author}{\bibfnamefont{H.}~\bibnamefont{Terai}},
  \bibinfo{author}{\bibfnamefont{R.}~\bibnamefont{Shimizu}},
  \bibinfo{author}{\bibfnamefont{M.}~\bibnamefont{Takeoka}}, \bibnamefont{and}
  \bibinfo{author}{\bibfnamefont{M.}~\bibnamefont{Sasaki}},
  \bibinfo{journal}{Opt. Express} \textbf{\bibinfo{volume}{23}},
  \bibinfo{pages}{28836} (\bibinfo{year}{2015}).

\bibitem[{\citenamefont{Prakash et~al.}()\citenamefont{Prakash, Sierant, and
  Mitchell}}]{Prakash2021}
\bibinfo{author}{\bibfnamefont{V.}~\bibnamefont{Prakash}},
  \bibinfo{author}{\bibfnamefont{A.}~\bibnamefont{Sierant}}, \bibnamefont{and}
  \bibinfo{author}{\bibfnamefont{M.~W.} \bibnamefont{Mitchell}},
  \bibinfo{howpublished}{arXiv:2101.03144}.

\bibitem[{\citenamefont{Montaut et~al.}(2018)\citenamefont{Montaut, {n}a
  Loaiza, Bartley, Verma, Nam, Mirin, Silberhorn, and Gerrits}}]{Montaut2018}
\bibinfo{author}{\bibfnamefont{N.}~\bibnamefont{Montaut}},
  \bibinfo{author}{\bibfnamefont{O.~S.~M.} \bibnamefont{{n}a Loaiza}},
  \bibinfo{author}{\bibfnamefont{T.~J.} \bibnamefont{Bartley}},
  \bibinfo{author}{\bibfnamefont{V.~B.} \bibnamefont{Verma}},
  \bibinfo{author}{\bibfnamefont{S.~W.} \bibnamefont{Nam}},
  \bibinfo{author}{\bibfnamefont{R.~P.} \bibnamefont{Mirin}},
  \bibinfo{author}{\bibfnamefont{C.}~\bibnamefont{Silberhorn}},
  \bibnamefont{and} \bibinfo{author}{\bibfnamefont{T.}~\bibnamefont{Gerrits}},
  \bibinfo{journal}{Optica} \textbf{\bibinfo{volume}{5}}, \bibinfo{pages}{1418}
  (\bibinfo{year}{2018}).

\bibitem[{\citenamefont{Gerrits et~al.}(2015)\citenamefont{Gerrits, Marsili,
  Verma, Shalm, Shaw, Mirin, and Nam}}]{Gerrits2015}
\bibinfo{author}{\bibfnamefont{T.}~\bibnamefont{Gerrits}},
  \bibinfo{author}{\bibfnamefont{F.}~\bibnamefont{Marsili}},
  \bibinfo{author}{\bibfnamefont{V.~B.} \bibnamefont{Verma}},
  \bibinfo{author}{\bibfnamefont{L.~K.} \bibnamefont{Shalm}},
  \bibinfo{author}{\bibfnamefont{M.}~\bibnamefont{Shaw}},
  \bibinfo{author}{\bibfnamefont{R.~P.} \bibnamefont{Mirin}}, \bibnamefont{and}
  \bibinfo{author}{\bibfnamefont{S.~W.} \bibnamefont{Nam}},
  \bibinfo{journal}{Phys. Rev. A} \textbf{\bibinfo{volume}{91}},
  \bibinfo{pages}{013830} (\bibinfo{year}{2015}).

\bibitem[{\citenamefont{Orre et~al.}(2019)\citenamefont{Orre, Goldschmidt,
  Deshpande, Gorshkov, Tamma, Hafezi, and Mittal}}]{Orre2019}
\bibinfo{author}{\bibfnamefont{V.~V.} \bibnamefont{Orre}},
  \bibinfo{author}{\bibfnamefont{E.~A.} \bibnamefont{Goldschmidt}},
  \bibinfo{author}{\bibfnamefont{A.}~\bibnamefont{Deshpande}},
  \bibinfo{author}{\bibfnamefont{A.~V.} \bibnamefont{Gorshkov}},
  \bibinfo{author}{\bibfnamefont{V.}~\bibnamefont{Tamma}},
  \bibinfo{author}{\bibfnamefont{M.}~\bibnamefont{Hafezi}}, \bibnamefont{and}
  \bibinfo{author}{\bibfnamefont{S.}~\bibnamefont{Mittal}},
  \bibinfo{journal}{Phys. Rev. Lett.} \textbf{\bibinfo{volume}{123}},
  \bibinfo{pages}{123603} (\bibinfo{year}{2019}).

\bibitem[{\citenamefont{Kaufman et~al.}(2014)\citenamefont{Kaufman, Lester,
  Reynolds, Wall, Foss-Feig, Hazzard, Rey, and Regal}}]{Kaufman2014}
\bibinfo{author}{\bibfnamefont{A.~M.} \bibnamefont{Kaufman}},
  \bibinfo{author}{\bibfnamefont{B.~J.} \bibnamefont{Lester}},
  \bibinfo{author}{\bibfnamefont{C.~M.} \bibnamefont{Reynolds}},
  \bibinfo{author}{\bibfnamefont{M.~L.} \bibnamefont{Wall}},
  \bibinfo{author}{\bibfnamefont{M.}~\bibnamefont{Foss-Feig}},
  \bibinfo{author}{\bibfnamefont{K.~R.~A.} \bibnamefont{Hazzard}},
  \bibinfo{author}{\bibfnamefont{A.~M.} \bibnamefont{Rey}}, \bibnamefont{and}
  \bibinfo{author}{\bibfnamefont{C.~A.} \bibnamefont{Regal}},
  \bibinfo{journal}{Science} \textbf{\bibinfo{volume}{345}},
  \bibinfo{pages}{306} (\bibinfo{year}{2014}).

\bibitem[{\citenamefont{Lopes et~al.}(2015)\citenamefont{Lopes, Imanaliev,
  Aspect, Cheneau, Boiron, and Westbrook}}]{Lopes2015}
\bibinfo{author}{\bibfnamefont{R.}~\bibnamefont{Lopes}},
  \bibinfo{author}{\bibfnamefont{A.}~\bibnamefont{Imanaliev}},
  \bibinfo{author}{\bibfnamefont{A.}~\bibnamefont{Aspect}},
  \bibinfo{author}{\bibfnamefont{M.}~\bibnamefont{Cheneau}},
  \bibinfo{author}{\bibfnamefont{D.}~\bibnamefont{Boiron}}, \bibnamefont{and}
  \bibinfo{author}{\bibfnamefont{C.~I.} \bibnamefont{Westbrook}},
  \bibinfo{journal}{Nature} \textbf{\bibinfo{volume}{520}}, \bibinfo{pages}{66}
  (\bibinfo{year}{2015}).

\bibitem[{\citenamefont{Toyoda et~al.}(2015)\citenamefont{Toyoda, Hiji,
  Noguchi, and Urabe}}]{Toyoda2015}
\bibinfo{author}{\bibfnamefont{K.}~\bibnamefont{Toyoda}},
  \bibinfo{author}{\bibfnamefont{R.}~\bibnamefont{Hiji}},
  \bibinfo{author}{\bibfnamefont{A.}~\bibnamefont{Noguchi}}, \bibnamefont{and}
  \bibinfo{author}{\bibfnamefont{S.}~\bibnamefont{Urabe}},
  \bibinfo{journal}{Nature} \textbf{\bibinfo{volume}{527}}, \bibinfo{pages}{74}
  (\bibinfo{year}{2015}).

\bibitem[{\citenamefont{Li et~al.}(2016)\citenamefont{Li, Zhou, Jing, Wang,
  Yang, Jiang, M\o{}lmer, Bao, and Pan}}]{Li2016}
\bibinfo{author}{\bibfnamefont{J.}~\bibnamefont{Li}},
  \bibinfo{author}{\bibfnamefont{M.-T.} \bibnamefont{Zhou}},
  \bibinfo{author}{\bibfnamefont{B.}~\bibnamefont{Jing}},
  \bibinfo{author}{\bibfnamefont{X.-J.} \bibnamefont{Wang}},
  \bibinfo{author}{\bibfnamefont{S.-J.} \bibnamefont{Yang}},
  \bibinfo{author}{\bibfnamefont{X.}~\bibnamefont{Jiang}},
  \bibinfo{author}{\bibfnamefont{K.}~\bibnamefont{M\o{}lmer}},
  \bibinfo{author}{\bibfnamefont{X.-H.} \bibnamefont{Bao}}, \bibnamefont{and}
  \bibinfo{author}{\bibfnamefont{J.-W.} \bibnamefont{Pan}},
  \bibinfo{journal}{Phys. Rev. Lett.} \textbf{\bibinfo{volume}{117}},
  \bibinfo{pages}{180501} (\bibinfo{year}{2016}).

\bibitem[{\citenamefont{Parniak et~al.}(2019)\citenamefont{Parniak, Mazelanik,
  Leszczy{\'{n}}ski, Lipka, D{\k{a}}browski, and Wasilewski}}]{Parniak2019}
\bibinfo{author}{\bibfnamefont{M.}~\bibnamefont{Parniak}},
  \bibinfo{author}{\bibfnamefont{M.}~\bibnamefont{Mazelanik}},
  \bibinfo{author}{\bibfnamefont{A.}~\bibnamefont{Leszczy{\'{n}}ski}},
  \bibinfo{author}{\bibfnamefont{M.}~\bibnamefont{Lipka}},
  \bibinfo{author}{\bibfnamefont{M.}~\bibnamefont{D{\k{a}}browski}},
  \bibnamefont{and}
  \bibinfo{author}{\bibfnamefont{W.}~\bibnamefont{Wasilewski}},
  \bibinfo{journal}{Phys. Rev. Lett.} \textbf{\bibinfo{volume}{122}},
  \bibinfo{pages}{063604} (\bibinfo{year}{2019}).

\bibitem[{\citenamefont{Uribe-Patarroyo
  et~al.}(2013)\citenamefont{Uribe-Patarroyo, Fraine, Simon, Minaeva, and
  Sergienko}}]{UribePatarroyo2013}
\bibinfo{author}{\bibfnamefont{N.}~\bibnamefont{Uribe-Patarroyo}},
  \bibinfo{author}{\bibfnamefont{A.}~\bibnamefont{Fraine}},
  \bibinfo{author}{\bibfnamefont{D.~S.} \bibnamefont{Simon}},
  \bibinfo{author}{\bibfnamefont{O.}~\bibnamefont{Minaeva}}, \bibnamefont{and}
  \bibinfo{author}{\bibfnamefont{A.~V.} \bibnamefont{Sergienko}},
  \bibinfo{journal}{Phys. Rev. Lett.} \textbf{\bibinfo{volume}{110}},
  \bibinfo{pages}{043601} (\bibinfo{year}{2013}).

\bibitem[{\citenamefont{Fickler et~al.}(2012)\citenamefont{Fickler,
  {\L}apkiewicz, Plick, Krenn, Schaeff, Ramelow, and Zeilinger}}]{Fickler2012}
\bibinfo{author}{\bibfnamefont{R.}~\bibnamefont{Fickler}},
  \bibinfo{author}{\bibfnamefont{R.}~\bibnamefont{{\L}apkiewicz}},
  \bibinfo{author}{\bibfnamefont{W.~N.} \bibnamefont{Plick}},
  \bibinfo{author}{\bibfnamefont{M.}~\bibnamefont{Krenn}},
  \bibinfo{author}{\bibfnamefont{C.}~\bibnamefont{Schaeff}},
  \bibinfo{author}{\bibfnamefont{S.}~\bibnamefont{Ramelow}}, \bibnamefont{and}
  \bibinfo{author}{\bibfnamefont{A.}~\bibnamefont{Zeilinger}},
  \bibinfo{journal}{Science} \textbf{\bibinfo{volume}{338}},
  \bibinfo{pages}{640} (\bibinfo{year}{2012}).

\bibitem[{\citenamefont{Leach et~al.}(2010)\citenamefont{Leach, Jack, Romero,
  Jha, Yao, Franke-Arnold, Ireland, Boyd, Barnett, and Padgett}}]{Leach2010}
\bibinfo{author}{\bibfnamefont{J.}~\bibnamefont{Leach}},
  \bibinfo{author}{\bibfnamefont{B.}~\bibnamefont{Jack}},
  \bibinfo{author}{\bibfnamefont{J.}~\bibnamefont{Romero}},
  \bibinfo{author}{\bibfnamefont{A.~K.} \bibnamefont{Jha}},
  \bibinfo{author}{\bibfnamefont{A.~M.} \bibnamefont{Yao}},
  \bibinfo{author}{\bibfnamefont{S.}~\bibnamefont{Franke-Arnold}},
  \bibinfo{author}{\bibfnamefont{D.~G.} \bibnamefont{Ireland}},
  \bibinfo{author}{\bibfnamefont{R.~W.} \bibnamefont{Boyd}},
  \bibinfo{author}{\bibfnamefont{S.~M.} \bibnamefont{Barnett}},
  \bibnamefont{and} \bibinfo{author}{\bibfnamefont{M.~J.}
  \bibnamefont{Padgett}}, \bibinfo{journal}{Science}
  \textbf{\bibinfo{volume}{329}}, \bibinfo{pages}{662} (\bibinfo{year}{2010}).

\bibitem[{\citenamefont{Karpi{\'{n}}ski
  et~al.}(2017)\citenamefont{Karpi{\'{n}}ski, Jachura, Wright, and
  Smith}}]{Karpinski2017}
\bibinfo{author}{\bibfnamefont{M.}~\bibnamefont{Karpi{\'{n}}ski}},
  \bibinfo{author}{\bibfnamefont{M.}~\bibnamefont{Jachura}},
  \bibinfo{author}{\bibfnamefont{L.~J.} \bibnamefont{Wright}},
  \bibnamefont{and} \bibinfo{author}{\bibfnamefont{B.~J.} \bibnamefont{Smith}},
  \bibinfo{journal}{Nat. Photonics} \textbf{\bibinfo{volume}{11}},
  \bibinfo{pages}{53} (\bibinfo{year}{2017}).

\bibitem[{\citenamefont{Kues et~al.}(2017)\citenamefont{Kues, Reimer, Roztocki,
  Cort\'{e}s, Sciara, Wetzel, Zhang, Cino, Chu, Little et~al.}}]{Kues2017}
\bibinfo{author}{\bibfnamefont{M.}~\bibnamefont{Kues}},
  \bibinfo{author}{\bibfnamefont{C.}~\bibnamefont{Reimer}},
  \bibinfo{author}{\bibfnamefont{P.}~\bibnamefont{Roztocki}},
  \bibinfo{author}{\bibfnamefont{L.~R.} \bibnamefont{Cort\'{e}s}},
  \bibinfo{author}{\bibfnamefont{S.}~\bibnamefont{Sciara}},
  \bibinfo{author}{\bibfnamefont{B.}~\bibnamefont{Wetzel}},
  \bibinfo{author}{\bibfnamefont{Y.}~\bibnamefont{Zhang}},
  \bibinfo{author}{\bibfnamefont{A.}~\bibnamefont{Cino}},
  \bibinfo{author}{\bibfnamefont{S.~T.} \bibnamefont{Chu}},
  \bibinfo{author}{\bibfnamefont{B.~E.} \bibnamefont{Little}},
  \bibnamefont{et~al.}, \bibinfo{journal}{Nature}
  \textbf{\bibinfo{volume}{546}}, \bibinfo{pages}{622} (\bibinfo{year}{2017}).

\bibitem[{\citenamefont{Nitsche et~al.}(2020)\citenamefont{Nitsche, De,
  Barkhofen, Meyer-Scott, Tiedau, Sperling, G\'{a}bris, Jex, and
  Silberhorn}}]{Nitsche2020}
\bibinfo{author}{\bibfnamefont{T.}~\bibnamefont{Nitsche}},
  \bibinfo{author}{\bibfnamefont{S.}~\bibnamefont{De}},
  \bibinfo{author}{\bibfnamefont{S.}~\bibnamefont{Barkhofen}},
  \bibinfo{author}{\bibfnamefont{E.}~\bibnamefont{Meyer-Scott}},
  \bibinfo{author}{\bibfnamefont{J.}~\bibnamefont{Tiedau}},
  \bibinfo{author}{\bibfnamefont{J.}~\bibnamefont{Sperling}},
  \bibinfo{author}{\bibfnamefont{A.}~\bibnamefont{G\'{a}bris}},
  \bibinfo{author}{\bibfnamefont{I.}~\bibnamefont{Jex}}, \bibnamefont{and}
  \bibinfo{author}{\bibfnamefont{C.}~\bibnamefont{Silberhorn}},
  \bibinfo{journal}{Phys. Rev. Lett.} \textbf{\bibinfo{volume}{125}},
  \bibinfo{pages}{213604} (\bibinfo{year}{2020}).

\bibitem[{\citenamefont{McCall and Hahn}(1967)}]{McCall1967}
\bibinfo{author}{\bibfnamefont{S.~L.} \bibnamefont{McCall}} \bibnamefont{and}
  \bibinfo{author}{\bibfnamefont{E.~L.} \bibnamefont{Hahn}},
  \bibinfo{journal}{Phys. Rev. Lett.} \textbf{\bibinfo{volume}{18}},
  \bibinfo{pages}{908} (\bibinfo{year}{1967}).

\bibitem[{\citenamefont{Crisp}(1970)}]{Crisp1970}
\bibinfo{author}{\bibfnamefont{M.~D.} \bibnamefont{Crisp}},
  \bibinfo{journal}{Phys. Rev. A} \textbf{\bibinfo{volume}{1}},
  \bibinfo{pages}{1604} (\bibinfo{year}{1970}).

\bibitem[{\citenamefont{Costanzo et~al.}(2016)\citenamefont{Costanzo, Coelho,
  Pellegrino, Mendes, Acioli, Cassemiro, Felinto, Zavatta, and
  Bellini}}]{Costanzo2016}
\bibinfo{author}{\bibfnamefont{L.~S.} \bibnamefont{Costanzo}},
  \bibinfo{author}{\bibfnamefont{A.~S.} \bibnamefont{Coelho}},
  \bibinfo{author}{\bibfnamefont{D.}~\bibnamefont{Pellegrino}},
  \bibinfo{author}{\bibfnamefont{M.~S.} \bibnamefont{Mendes}},
  \bibinfo{author}{\bibfnamefont{L.}~\bibnamefont{Acioli}},
  \bibinfo{author}{\bibfnamefont{K.~N.} \bibnamefont{Cassemiro}},
  \bibinfo{author}{\bibfnamefont{D.}~\bibnamefont{Felinto}},
  \bibinfo{author}{\bibfnamefont{A.}~\bibnamefont{Zavatta}}, \bibnamefont{and}
  \bibinfo{author}{\bibfnamefont{M.}~\bibnamefont{Bellini}},
  \bibinfo{journal}{Phys. Rev. Lett.} \textbf{\bibinfo{volume}{116}},
  \bibinfo{pages}{023602} (\bibinfo{year}{2016}).

\bibitem[{\citenamefont{Meshulach and Silberberg}(1998)}]{Meshulach1998}
\bibinfo{author}{\bibfnamefont{D.}~\bibnamefont{Meshulach}} \bibnamefont{and}
  \bibinfo{author}{\bibfnamefont{Y.}~\bibnamefont{Silberberg}},
  \bibinfo{journal}{Nature} \textbf{\bibinfo{volume}{396}},
  \bibinfo{pages}{239} (\bibinfo{year}{1998}).

\bibitem[{\citenamefont{Devolder et~al.}(2021)\citenamefont{Devolder, Brumer,
  and Tscherbul}}]{Devolder2021}
\bibinfo{author}{\bibfnamefont{A.}~\bibnamefont{Devolder}},
  \bibinfo{author}{\bibfnamefont{P.}~\bibnamefont{Brumer}}, \bibnamefont{and}
  \bibinfo{author}{\bibfnamefont{T.~V.} \bibnamefont{Tscherbul}},
  \bibinfo{journal}{Phys. Rev. Lett.} \textbf{\bibinfo{volume}{126}},
  \bibinfo{pages}{153403} (\bibinfo{year}{2021}).

\bibitem[{\citenamefont{Carvalho et~al.}(2020)\citenamefont{Carvalho, Moreira,
  Ferraz, Vianna, Acioli, and Felinto}}]{Carvalho2020}
\bibinfo{author}{\bibfnamefont{A.~J.~A.} \bibnamefont{Carvalho}},
  \bibinfo{author}{\bibfnamefont{R.~S.~N.} \bibnamefont{Moreira}},
  \bibinfo{author}{\bibfnamefont{J.}~\bibnamefont{Ferraz}},
  \bibinfo{author}{\bibfnamefont{S.~S.} \bibnamefont{Vianna}},
  \bibinfo{author}{\bibfnamefont{L.~H.} \bibnamefont{Acioli}},
  \bibnamefont{and} \bibinfo{author}{\bibfnamefont{D.}~\bibnamefont{Felinto}},
  \bibinfo{journal}{Phys. Rev. A} \textbf{\bibinfo{volume}{101}},
  \bibinfo{pages}{053426} (\bibinfo{year}{2020}).

\bibitem[{\citenamefont{Marian et~al.}(2004)\citenamefont{Marian, Stowe,
  Lawall, Felinto, and Ye}}]{Marian2004}
\bibinfo{author}{\bibfnamefont{A.}~\bibnamefont{Marian}},
  \bibinfo{author}{\bibfnamefont{M.~C.} \bibnamefont{Stowe}},
  \bibinfo{author}{\bibfnamefont{J.~R.} \bibnamefont{Lawall}},
  \bibinfo{author}{\bibfnamefont{D.}~\bibnamefont{Felinto}}, \bibnamefont{and}
  \bibinfo{author}{\bibfnamefont{J.}~\bibnamefont{Ye}},
  \bibinfo{journal}{Science} \textbf{\bibinfo{volume}{306}},
  \bibinfo{pages}{2063} (\bibinfo{year}{2004}).

\bibitem[{\citenamefont{Polycarpou et~al.}(2012)\citenamefont{Polycarpou,
  Cassemiro, Venturi, Zavatta, and Bellini}}]{Polycarpou2012}
\bibinfo{author}{\bibfnamefont{C.}~\bibnamefont{Polycarpou}},
  \bibinfo{author}{\bibfnamefont{K.~N.} \bibnamefont{Cassemiro}},
  \bibinfo{author}{\bibfnamefont{G.}~\bibnamefont{Venturi}},
  \bibinfo{author}{\bibfnamefont{A.}~\bibnamefont{Zavatta}}, \bibnamefont{and}
  \bibinfo{author}{\bibfnamefont{M.}~\bibnamefont{Bellini}},
  \bibinfo{journal}{Phys. Rev. Lett.} \textbf{\bibinfo{volume}{109}},
  \bibinfo{pages}{053602} (\bibinfo{year}{2012}).

\bibitem[{SM()}]{SM}
\bibinfo{howpublished}{See Supplemental Material for additional
  calculations and derivations, details on the role of photons' spectral
  correlations, phase reconstruction strategy, visibility estimates, comparison
  with homodyne tomography, and derivation of Fisher information for estimating
  the residual distinguishability, which includes Refs. [33-44].}

\bibitem[{\citenamefont{Nesmeyanov and Gary}(1963)}]{nesmeyanov_gary_1963}
\bibinfo{author}{\bibfnamefont{A.}~\bibnamefont{Nesmeyanov}} \bibnamefont{and}
  \bibinfo{author}{\bibfnamefont{R.}~\bibnamefont{Gary}},
  \emph{\bibinfo{title}{Vapor pressure of the chemical elements}}
  (\bibinfo{publisher}{Elsevier}, \bibinfo{year}{1963}).

\bibitem[{\citenamefont{Kallmann et~al.}(1999)\citenamefont{Kallmann, Brattke,
  and Hartmann}}]{Kallmann1999}
\bibinfo{author}{\bibfnamefont{U.}~\bibnamefont{Kallmann}},
  \bibinfo{author}{\bibfnamefont{S.}~\bibnamefont{Brattke}}, \bibnamefont{and}
  \bibinfo{author}{\bibfnamefont{W.}~\bibnamefont{Hartmann}},
  \bibinfo{journal}{Phys. Rev. A} \textbf{\bibinfo{volume}{59}},
  \bibinfo{pages}{814} (\bibinfo{year}{1999}).

\bibitem[{\citenamefont{Kolenderski et~al.}(2009)\citenamefont{Kolenderski,
  Wasilewski, and Banaszek}}]{Kolenderski2009}
\bibinfo{author}{\bibfnamefont{P.}~\bibnamefont{Kolenderski}},
  \bibinfo{author}{\bibfnamefont{W.}~\bibnamefont{Wasilewski}},
  \bibnamefont{and} \bibinfo{author}{\bibfnamefont{K.}~\bibnamefont{Banaszek}},
  \bibinfo{journal}{Phys. Rev. A} \textbf{\bibinfo{volume}{80}},
  \bibinfo{pages}{013811} (\bibinfo{year}{2009}).

\bibitem[{\citenamefont{Karout et~al.}(2007)\citenamefont{Karout, Gdeisat,
  Burton, and Lalor}}]{Karout2007}
\bibinfo{author}{\bibfnamefont{S.~A.} \bibnamefont{Karout}},
  \bibinfo{author}{\bibfnamefont{M.~A.} \bibnamefont{Gdeisat}},
  \bibinfo{author}{\bibfnamefont{D.~R.} \bibnamefont{Burton}},
  \bibnamefont{and} \bibinfo{author}{\bibfnamefont{M.~J.} \bibnamefont{Lalor}},
  \bibinfo{journal}{Appl. Opt.} \textbf{\bibinfo{volume}{46}},
  \bibinfo{pages}{730} (\bibinfo{year}{2007}).

\bibitem[{\citenamefont{Bone et~al.}(1986)\citenamefont{Bone, Bachor, and
  Sandeman}}]{Bone1986}
\bibinfo{author}{\bibfnamefont{D.~J.} \bibnamefont{Bone}},
  \bibinfo{author}{\bibfnamefont{H.-A.} \bibnamefont{Bachor}},
  \bibnamefont{and} \bibinfo{author}{\bibfnamefont{R.~J.}
  \bibnamefont{Sandeman}}, \bibinfo{journal}{Appl. Opt.}
  \textbf{\bibinfo{volume}{25}}, \bibinfo{pages}{1653} (\bibinfo{year}{1986}).

\bibitem[{\citenamefont{Kay}(1993)}]{Kay1993}
\bibinfo{author}{\bibfnamefont{S.~M.} \bibnamefont{Kay}},
  \emph{\bibinfo{title}{{Fundamentals of Statistical Signal Processing:
  Estimation Theory}}} (\bibinfo{publisher}{PTR Prentice Hall},
  \bibinfo{year}{1993}).

\bibitem[{\citenamefont{Sekatski et~al.}(2012)\citenamefont{Sekatski,
  Sangouard, Bussi\`{e}res, Clausen, Gisin, and Zbinden}}]{Sekatski2012}
\bibinfo{author}{\bibfnamefont{P.}~\bibnamefont{Sekatski}},
  \bibinfo{author}{\bibfnamefont{N.}~\bibnamefont{Sangouard}},
  \bibinfo{author}{\bibfnamefont{F.}~\bibnamefont{Bussi\`{e}res}},
  \bibinfo{author}{\bibfnamefont{C.}~\bibnamefont{Clausen}},
  \bibinfo{author}{\bibfnamefont{N.}~\bibnamefont{Gisin}}, \bibnamefont{and}
  \bibinfo{author}{\bibfnamefont{H.}~\bibnamefont{Zbinden}},
  \bibinfo{journal}{{J. Phys. B: At. Mol. Opt. Phys.}}
  \textbf{\bibinfo{volume}{45}}, \bibinfo{pages}{124016}
  (\bibinfo{year}{2012}).

\bibitem[{\citenamefont{Schleich}(2001)}]{Schleich2001}
\bibinfo{author}{\bibfnamefont{W.~P.} \bibnamefont{Schleich}},
  \emph{\bibinfo{title}{Quantum Optics in Phase Space}}
  (\bibinfo{publisher}{Wiley-VCH}, \bibinfo{year}{2001}), ISBN
  \bibinfo{isbn}{978-3-527-29435-0}.

\bibitem[{\citenamefont{Garrison and Chiao}(2008)}]{GarrisonChiao}
\bibinfo{author}{\bibfnamefont{J.~C.} \bibnamefont{Garrison}} \bibnamefont{and}
  \bibinfo{author}{\bibfnamefont{R.~Y.} \bibnamefont{Chiao}},
  \emph{\bibinfo{title}{Quantum Optics}} (\bibinfo{publisher}{Oxford University
  Press Inc.}, \bibinfo{year}{2008}), ISBN \bibinfo{isbn}{9780198508861}.

\bibitem[{\citenamefont{Henriksson et~al.}(2009)\citenamefont{Henriksson,
  McDermott, and Bergmanson}}]{Henriksson2009}
\bibinfo{author}{\bibfnamefont{J.~T.} \bibnamefont{Henriksson}},
  \bibinfo{author}{\bibfnamefont{A.~M.} \bibnamefont{McDermott}},
  \bibnamefont{and} \bibinfo{author}{\bibfnamefont{J.~P.~G.}
  \bibnamefont{Bergmanson}}, \bibinfo{journal}{{Invest. Ophthalmol. Vis. Sci.}}
  \textbf{\bibinfo{volume}{50}}, \bibinfo{pages}{3648} (\bibinfo{year}{2009}).

\bibitem[{\citenamefont{Tang et~al.}(2019)\citenamefont{Tang, Liu, Chen, Yu,
  Luo, Wu, Wang, and Liu}}]{Tang2019}
\bibinfo{author}{\bibfnamefont{H.}~\bibnamefont{Tang}},
  \bibinfo{author}{\bibfnamefont{X.}~\bibnamefont{Liu}},
  \bibinfo{author}{\bibfnamefont{S.}~\bibnamefont{Chen}},
  \bibinfo{author}{\bibfnamefont{X.}~\bibnamefont{Yu}},
  \bibinfo{author}{\bibfnamefont{Y.}~\bibnamefont{Luo}},
  \bibinfo{author}{\bibfnamefont{J.}~\bibnamefont{Wu}},
  \bibinfo{author}{\bibfnamefont{X.}~\bibnamefont{Wang}}, \bibnamefont{and}
  \bibinfo{author}{\bibfnamefont{L.}~\bibnamefont{Liu}},
  \bibinfo{journal}{{IEEE. Trans. Biomed}} \textbf{\bibinfo{volume}{66}},
  \bibinfo{pages}{1803} (\bibinfo{year}{2019}).

\bibitem[{\citenamefont{Jacques}(2013)}]{Jacques2013}
\bibinfo{author}{\bibfnamefont{S.~L.} \bibnamefont{Jacques}},
  \bibinfo{journal}{{Phys. Med. Biol}} \textbf{\bibinfo{volume}{58}},
  \bibinfo{pages}{R37} (\bibinfo{year}{2013}).

\bibitem[{\citenamefont{Wallauer et~al.}(2021)\citenamefont{Wallauer, Raths,
  Stallberg, M\"{u}nster, Brandstetter, Yang, G\"{u}dde, Puschnig, Soubatch,
  Kumpf et~al.}}]{Wallauer2021}
\bibinfo{author}{\bibfnamefont{R.}~\bibnamefont{Wallauer}},
  \bibinfo{author}{\bibfnamefont{M.}~\bibnamefont{Raths}},
  \bibinfo{author}{\bibfnamefont{K.}~\bibnamefont{Stallberg}},
  \bibinfo{author}{\bibfnamefont{L.}~\bibnamefont{M\"{u}nster}},
  \bibinfo{author}{\bibfnamefont{D.}~\bibnamefont{Brandstetter}},
  \bibinfo{author}{\bibfnamefont{X.}~\bibnamefont{Yang}},
  \bibinfo{author}{\bibfnamefont{J.}~\bibnamefont{G\"{u}dde}},
  \bibinfo{author}{\bibfnamefont{P.}~\bibnamefont{Puschnig}},
  \bibinfo{author}{\bibfnamefont{S.}~\bibnamefont{Soubatch}},
  \bibinfo{author}{\bibfnamefont{C.}~\bibnamefont{Kumpf}},
  \bibnamefont{et~al.}, \bibinfo{journal}{Science}
  \textbf{\bibinfo{volume}{371}}, \bibinfo{pages}{1056} (\bibinfo{year}{2021}).

\bibitem[{\citenamefont{Rosker et~al.}(1988)\citenamefont{Rosker, Dantus, and
  Zewail}}]{ROSKER1988}
\bibinfo{author}{\bibfnamefont{M.~J.} \bibnamefont{Rosker}},
  \bibinfo{author}{\bibfnamefont{M.}~\bibnamefont{Dantus}}, \bibnamefont{and}
  \bibinfo{author}{\bibfnamefont{A.~H.} \bibnamefont{Zewail}},
  \bibinfo{journal}{Science} \textbf{\bibinfo{volume}{241}},
  \bibinfo{pages}{1200} (\bibinfo{year}{1988}).

\bibitem[{\citenamefont{Dantus et~al.}(1987)\citenamefont{Dantus, Rosker, and
  Zewail}}]{Dantus1987}
\bibinfo{author}{\bibfnamefont{M.}~\bibnamefont{Dantus}},
  \bibinfo{author}{\bibfnamefont{M.~J.} \bibnamefont{Rosker}},
  \bibnamefont{and} \bibinfo{author}{\bibfnamefont{A.~H.}
  \bibnamefont{Zewail}}, \bibinfo{journal}{J. Chem. Phys.}
  \textbf{\bibinfo{volume}{87}}, \bibinfo{pages}{2395} (\bibinfo{year}{1987}).

\bibitem[{\citenamefont{Potter et~al.}(1992)\citenamefont{Potter, Herek,
  Pedersen, Liu, and Zewail}}]{Potter1992}
\bibinfo{author}{\bibfnamefont{E.~D.} \bibnamefont{Potter}},
  \bibinfo{author}{\bibfnamefont{J.~L.} \bibnamefont{Herek}},
  \bibinfo{author}{\bibfnamefont{S.}~\bibnamefont{Pedersen}},
  \bibinfo{author}{\bibfnamefont{Q.}~\bibnamefont{Liu}}, \bibnamefont{and}
  \bibinfo{author}{\bibfnamefont{A.~H.} \bibnamefont{Zewail}},
  \bibinfo{journal}{Nature} \textbf{\bibinfo{volume}{355}}, \bibinfo{pages}{66}
  (\bibinfo{year}{1992}).

\bibitem[{\citenamefont{Brumer and Shapiro}(1992)}]{Brumer1992}
\bibinfo{author}{\bibfnamefont{P.}~\bibnamefont{Brumer}} \bibnamefont{and}
  \bibinfo{author}{\bibfnamefont{M.}~\bibnamefont{Shapiro}},
  \bibinfo{journal}{Annu. Rev. Phys. Chem.} \textbf{\bibinfo{volume}{43}},
  \bibinfo{pages}{257} (\bibinfo{year}{1992}).

\bibitem[{\citenamefont{Mukamel et~al.}(2020)\citenamefont{Mukamel, Freyberger,
  Schleich, Bellini, Zavatta, Leuchs, Silberhorn, Boyd, S\'{a}nchez-Soto,
  Stefanov et~al.}}]{Mukamel2020}
\bibinfo{author}{\bibfnamefont{S.}~\bibnamefont{Mukamel}},
  \bibinfo{author}{\bibfnamefont{M.}~\bibnamefont{Freyberger}},
  \bibinfo{author}{\bibfnamefont{W.}~\bibnamefont{Schleich}},
  \bibinfo{author}{\bibfnamefont{M.}~\bibnamefont{Bellini}},
  \bibinfo{author}{\bibfnamefont{A.}~\bibnamefont{Zavatta}},
  \bibinfo{author}{\bibfnamefont{G.}~\bibnamefont{Leuchs}},
  \bibinfo{author}{\bibfnamefont{C.}~\bibnamefont{Silberhorn}},
  \bibinfo{author}{\bibfnamefont{R.~W.} \bibnamefont{Boyd}},
  \bibinfo{author}{\bibfnamefont{L.~L.} \bibnamefont{S\'{a}nchez-Soto}},
  \bibinfo{author}{\bibfnamefont{A.}~\bibnamefont{Stefanov}},
  \bibnamefont{et~al.}, \bibinfo{journal}{Journal of Physics B: Atomic,
  Molecular and Optical Physics} \textbf{\bibinfo{volume}{53}},
  \bibinfo{pages}{072002} (\bibinfo{year}{2020}).

\bibitem[{\citenamefont{Dorfman et~al.}(2016)\citenamefont{Dorfman, Schlawin,
  and Mukamel}}]{Dorfman2016}
\bibinfo{author}{\bibfnamefont{K.~E.} \bibnamefont{Dorfman}},
  \bibinfo{author}{\bibfnamefont{F.}~\bibnamefont{Schlawin}}, \bibnamefont{and}
  \bibinfo{author}{\bibfnamefont{S.}~\bibnamefont{Mukamel}},
  \bibinfo{journal}{Rev. Mod. Phys.} \textbf{\bibinfo{volume}{88}},
  \bibinfo{pages}{045008} (\bibinfo{year}{2016}).

\bibitem[{\citenamefont{Dorfman et~al.}(2021)\citenamefont{Dorfman, Asban, Gu,
  and Mukamel}}]{Dorfman2021}
\bibinfo{author}{\bibfnamefont{K.~E.} \bibnamefont{Dorfman}},
  \bibinfo{author}{\bibfnamefont{S.}~\bibnamefont{Asban}},
  \bibinfo{author}{\bibfnamefont{B.}~\bibnamefont{Gu}}, \bibnamefont{and}
  \bibinfo{author}{\bibfnamefont{S.}~\bibnamefont{Mukamel}},
  \bibinfo{journal}{Communications Physics} \textbf{\bibinfo{volume}{4}},
  \bibinfo{pages}{49} (\bibinfo{year}{2021}).

\bibitem[{\citenamefont{Rose et~al.}(1988)\citenamefont{Rose, Rosker, and
  Zewail}}]{Rose1988}
\bibinfo{author}{\bibfnamefont{T.~S.} \bibnamefont{Rose}},
  \bibinfo{author}{\bibfnamefont{M.~J.} \bibnamefont{Rosker}},
  \bibnamefont{and} \bibinfo{author}{\bibfnamefont{A.~H.}
  \bibnamefont{Zewail}}, \bibinfo{journal}{J. Chem. Phys.}
  \textbf{\bibinfo{volume}{88}}, \bibinfo{pages}{6672} (\bibinfo{year}{1988}).

\bibitem[{\citenamefont{Brown et~al.}(2006)\citenamefont{Brown, Dicks, and
  Walmsley}}]{Brown2006}
\bibinfo{author}{\bibfnamefont{B.~L.} \bibnamefont{Brown}},
  \bibinfo{author}{\bibfnamefont{A.~J.} \bibnamefont{Dicks}}, \bibnamefont{and}
  \bibinfo{author}{\bibfnamefont{I.~A.} \bibnamefont{Walmsley}},
  \bibinfo{journal}{PRL} \textbf{\bibinfo{volume}{96}}, \bibinfo{pages}{173002}
  (\bibinfo{year}{2006}).

\bibitem[{\citenamefont{Meng et~al.}(2020)\citenamefont{Meng, Brawley, Bennett,
  Vanner, and Bowen}}]{Meng2020}
\bibinfo{author}{\bibfnamefont{C.}~\bibnamefont{Meng}},
  \bibinfo{author}{\bibfnamefont{G.~A.} \bibnamefont{Brawley}},
  \bibinfo{author}{\bibfnamefont{J.~S.} \bibnamefont{Bennett}},
  \bibinfo{author}{\bibfnamefont{M.~R.} \bibnamefont{Vanner}},
  \bibnamefont{and} \bibinfo{author}{\bibfnamefont{W.~P.} \bibnamefont{Bowen}},
  \bibinfo{journal}{PRL} \textbf{\bibinfo{volume}{125}},
  \bibinfo{pages}{043604} (\bibinfo{year}{2020}).

\bibitem[{\citenamefont{Klinteberg et~al.}(2005)\citenamefont{Klinteberg,
  Pifferi, Andersson-Engels, Cubeddu, and Svanberg}}]{Klinteberg2005}
\bibinfo{author}{\bibfnamefont{C.~a.} \bibnamefont{Klinteberg}},
  \bibinfo{author}{\bibfnamefont{A.}~\bibnamefont{Pifferi}},
  \bibinfo{author}{\bibfnamefont{S.}~\bibnamefont{Andersson-Engels}},
  \bibinfo{author}{\bibfnamefont{R.}~\bibnamefont{Cubeddu}}, \bibnamefont{and}
  \bibinfo{author}{\bibfnamefont{S.}~\bibnamefont{Svanberg}},
  \bibinfo{journal}{Appl. Opt.} \textbf{\bibinfo{volume}{44}},
  \bibinfo{pages}{2213} (\bibinfo{year}{2005}).

\bibitem[{\citenamefont{Hori et~al.}(2006)\citenamefont{Hori, Yasui, and
  Araki}}]{Hori2006}
\bibinfo{author}{\bibfnamefont{Y.}~\bibnamefont{Hori}},
  \bibinfo{author}{\bibfnamefont{T.}~\bibnamefont{Yasui}}, \bibnamefont{and}
  \bibinfo{author}{\bibfnamefont{T.}~\bibnamefont{Araki}},
  \bibinfo{journal}{Optical Review} \textbf{\bibinfo{volume}{13}},
  \bibinfo{pages}{29} (\bibinfo{year}{2006}).

\bibitem[{\citenamefont{Wolfgramm et~al.}(2013)\citenamefont{Wolfgramm,
  Vitelli, Beduini, Godbout, and Mitchell}}]{Wolfgramm2013}
\bibinfo{author}{\bibfnamefont{F.}~\bibnamefont{Wolfgramm}},
  \bibinfo{author}{\bibfnamefont{C.}~\bibnamefont{Vitelli}},
  \bibinfo{author}{\bibfnamefont{F.~A.} \bibnamefont{Beduini}},
  \bibinfo{author}{\bibfnamefont{N.}~\bibnamefont{Godbout}}, \bibnamefont{and}
  \bibinfo{author}{\bibfnamefont{M.~W.} \bibnamefont{Mitchell}},
  \bibinfo{journal}{Nature Photonics} \textbf{\bibinfo{volume}{7}},
  \bibinfo{pages}{28} (\bibinfo{year}{2013}).

\bibitem[{\citenamefont{Lipka and Parniak}(2021)}]{Lipka2021}
\bibinfo{author}{\bibfnamefont{M.}~\bibnamefont{Lipka}} \bibnamefont{and}
  \bibinfo{author}{\bibfnamefont{M.}~\bibnamefont{Parniak}},
  \bibinfo{journal}{Opt. Lett.} \textbf{\bibinfo{volume}{46}},
  \bibinfo{pages}{3009} (\bibinfo{year}{2021}).

\bibitem[{\citenamefont{Lipka et~al.}(2018)\citenamefont{Lipka, Parniak, and
  Wasilewski}}]{Lipka2018}
\bibinfo{author}{\bibfnamefont{M.}~\bibnamefont{Lipka}},
  \bibinfo{author}{\bibfnamefont{M.}~\bibnamefont{Parniak}}, \bibnamefont{and}
  \bibinfo{author}{\bibfnamefont{W.}~\bibnamefont{Wasilewski}},
  \bibinfo{journal}{Appl. Phys. Lett.} \textbf{\bibinfo{volume}{112}},
  \bibinfo{pages}{211105} (\bibinfo{year}{2018}).

\bibitem[{\citenamefont{Jozsa}(1994)}]{Jozsa1994}
\bibinfo{author}{\bibfnamefont{R.}~\bibnamefont{Jozsa}}, \bibinfo{journal}{J.
  Mod. Opt} \textbf{\bibinfo{volume}{41}}, \bibinfo{pages}{2315}
  (\bibinfo{year}{1994}).

\bibitem[{\citenamefont{Qin et~al.}(2015)\citenamefont{Qin, Prasad, Brannan,
  MacRae, Lezama, and Lvovsky}}]{Qin2015}
\bibinfo{author}{\bibfnamefont{Z.}~\bibnamefont{Qin}},
  \bibinfo{author}{\bibfnamefont{A.~S.} \bibnamefont{Prasad}},
  \bibinfo{author}{\bibfnamefont{T.}~\bibnamefont{Brannan}},
  \bibinfo{author}{\bibfnamefont{A.}~\bibnamefont{MacRae}},
  \bibinfo{author}{\bibfnamefont{A.}~\bibnamefont{Lezama}}, \bibnamefont{and}
  \bibinfo{author}{\bibfnamefont{A.~I.} \bibnamefont{Lvovsky}},
  \bibinfo{journal}{Light: Science {\&} Applications}
  \textbf{\bibinfo{volume}{4}}, \bibinfo{pages}{e298} (\bibinfo{year}{2015}).

\bibitem[{\citenamefont{Ferrie}(2014)}]{Ferrie2014}
\bibinfo{author}{\bibfnamefont{C.}~\bibnamefont{Ferrie}},
  \bibinfo{journal}{Phys. Rev. Lett.} \textbf{\bibinfo{volume}{113}},
  \bibinfo{pages}{190404} (\bibinfo{year}{2014}).

\bibitem[{\citenamefont{Brecht et~al.}(2015)\citenamefont{Brecht, Reddy,
  Silberhorn, and Raymer}}]{Brecht2015}
\bibinfo{author}{\bibfnamefont{B.}~\bibnamefont{Brecht}},
  \bibinfo{author}{\bibfnamefont{D.~V.} \bibnamefont{Reddy}},
  \bibinfo{author}{\bibfnamefont{C.}~\bibnamefont{Silberhorn}},
  \bibnamefont{and} \bibinfo{author}{\bibfnamefont{M.~G.}
  \bibnamefont{Raymer}}, \bibinfo{journal}{Phys. Rev. X}
  \textbf{\bibinfo{volume}{5}}, \bibinfo{pages}{041017} (\bibinfo{year}{2015}).

\bibitem[{\citenamefont{Lukens and Lougovski}(2017)}]{Lukens2017}
\bibinfo{author}{\bibfnamefont{J.~M.} \bibnamefont{Lukens}} \bibnamefont{and}
  \bibinfo{author}{\bibfnamefont{P.}~\bibnamefont{Lougovski}},
  \bibinfo{journal}{Optica} \textbf{\bibinfo{volume}{4}}, \bibinfo{pages}{8}
  (\bibinfo{year}{2017}).

\bibitem[{\citenamefont{Khodadad~Kashi and Kues}(2021)}]{KhodadadKashi2021}
\bibinfo{author}{\bibfnamefont{A.}~\bibnamefont{Khodadad~Kashi}}
  \bibnamefont{and} \bibinfo{author}{\bibfnamefont{M.}~\bibnamefont{Kues}},
  \bibinfo{journal}{Laser Photonics Rev.} \textbf{\bibinfo{volume}{15}},
  \bibinfo{pages}{2000464} (\bibinfo{year}{2021}).

\end{thebibliography}

\end{document}